\documentclass[apj]{emulateapj}

\begin{document}

\title{The $\sigma - L$ correlation in Nearby Early-Type Galaxies}

\author{Mariangela Bernardi\altaffilmark{1}}

\affil{Dept. of Physics and Astronomy, University of Pennsylvania, 
                 209 South 33rd St, Philadelphia, PA 19104}

\altaffiltext{1}{bernardm@physics.upenn.edu}

\begin{abstract}
Early-type galaxy velocity dispersions and luminosities are 
correlated.  The correlation estimated in local samples ($\le 100$~Mpc) 
differs from that measured more recently in the SDSS.
This is true even when systematics in the SDSS photometric and spectroscopic
parameters have been accounted-for.
We show that this is also true for the ENEAR sample if galaxy luminosities are 
estimated using distances which have been corrected for peculiar 
motions.  We then show that, because the estimate of the `true' 
distance is derived from a correlation with velocity dispersion, 
in this case the $D_n-\sigma$ relation, using it in the $\sigma-L$ 
relation leads to an artificially tight relation with a biased slope.  
Making no correction for peculiar velocities results in a $\sigma-L$ 
relation which is very similar to that of the SDSS, although with 
larger scatter.  We also measure the $\sigma-L$ correlation in a mock 
ENEAR catalog, in which the underlying galaxy sample has the same 
$\sigma-L$ correlation as seen in the SDSS.  The mock catalog 
produces the same $D_n-\sigma$ relation as the data, the same biased 
slope when $D_n-\sigma$ distances are used to estimate luminosities, 
and good agreement with the input $\sigma-L$ relation when redshift 
is used as the distance indicator.  
This provides further evidence that the true $\sigma-L$ relation of 
ENEAR galaxies is indeed very similar to that of SDSS early-types.  
Our results suggest that local $\sigma-L$ relations which are based 
on Fundamental Plane distances should also be re-evaluated.  
%Although we report the results of fitting a single power-law to 
%the $\sigma-L$ relation, we see evidence for curvature:  
%the velocity dispersions of the most luminous galaxies are smaller 
%than predicted by the power-law fit.  
Our findings also have important implications for black hole 
demographics; the best direct estimates of the masses of supermassive 
black holes come from local galaxies, so estimates of the black 
hole mass function are more safely made by working with the 
$M_\bullet-\sigma$ correlation than with $M_\bullet-L$.  
\end{abstract}


\section{Introduction}
The luminosities and velocity dispersions of early-type galaxies 
are strongly correlated:  their logarithms follow an approximately
linear relation (e.g. Faber \& Jackson 1976).  
There is a long and complicated history of what the slope of this 
relation is, mainly due to the fact that, if there is intrinsic 
scatter around this correlation, then there are at least three 
interesting slopes:  that obtained from fitting the mean luminosity 
at each velocity dispersion, 
 $\langle\log L|\log\sigma\rangle = a_{L|\sigma}\log\sigma + b_{L|\sigma}$, 
the mean velocity dispersion as a function of luminosity, 
 $\langle\log\sigma|\log L\rangle = a_{\sigma|L}\log L + b_{\sigma|L}$, 
and the slope of the principal axis of the joint distribution of $L$ 
and $\sigma$ (e.g. Lynden-Bell et al. 1988; Saglia et al. 2001).  
The first of these allows one to use $\sigma$ to predict $L$, 
whereas the second must be used if one wishes to predict $\sigma$ 
from $L$.  Until recently, authors were not careful to distinguish 
between these cases.  

Because $L$ depends on the distance to the source, whereas $\sigma$ 
does not, it is straightforward to estimate 
 $\langle\log\sigma|\log L\rangle$ 
from flux limited samples; naive estimates of the other two 
correlations are compromised by selection effects.  
Most studies of this correlation, based on local samples, agree 
that the typical velocity dispersion at fixed luminosity, 
$\langle\log\sigma|\log L\rangle$, scales as $L^{1/4}$:  
 $a_{\sigma|L}\approx 1/4$.  
The common unfortunate abuse of jargon is to say that luminosity 
scales as the fourth power of velocity dispersion; in fact, 
$a_{L|\sigma}$ is considerably shallower.  
In what follows, we will be almost exclusively concerned with how 
well $L$ can be used as a predictor for $\sigma$:
  $\langle\log\sigma|\log L\rangle$.  

Prior to the year 2000, studies of the $\sigma-L$ correlation were 
restricted to local samples (from within about 50 Mpc) containing 
$\sim 100$ objects.  The SDSS provided a sample of early-type 
galaxies that was larger by about two orders of magnitude 
(Bernardi et al. 2003a):  $\sim 9000$ objects drawn from a volume 
which extended to considerably larger distances (median redshift 
$z\sim 0.1$).  However, the $\sigma-L$ relation in this sample,
\begin{equation}
 \Bigl\langle\log_{10}\sigma|M_r\Bigr\rangle_{\rm SDSS-B03} 
 = 2.203 - 0.102\,(M_r+21),
 \label{sdss03}
\end{equation}
with an intrinsic scatter of about 0.07~dex is inconsistent with 
that found in local samples.  
For example, at $\log_{10}(\sigma/{\rm km~s}^{-1}) \ge 2.4$, 
the relation which Forbes \& Ponman (1999) found best fits the 
Pruniel \& Simien (1996) sample of 236 local early-type galaxies 
differs from the SDSS fit by more than 0.05~dex.  
Expressed in terms of absolute magnitudes, the fits differ by more 
than 0.5~mags at $M_r\le -23$.  This is substantially larger than 
expected given the measurement errors.  What causes this difference?  

Because the SDSS luminosities and velocity dispersions are obtained 
from an automated pipeline, (i.e., light profiles and spectra were 
not inspected individually), it may be that, say, the SDSS measurements 
differ systematically from those estimated for local objects in the 
literature.  Indeed, it is known that the SDSS photometric 
reductions underestimate the luminosities of objects in crowded 
fields (Mandelbaum et al. 2005; Bernardi et al. 2006a; Hyde et al. 2006; 
Lauer et al. 2006). See the Appendix for a more detailed discussion.
The Appendix also shows that the SDSS also slightly overestimates the 
velocity dispersions at small $\sigma$.  
In what follows, we will refer to the sample in which these 
systematics have been accounted-for as the SDSS-B06 sample 
(see the Appendix).  
In the SDSS-B06 sample, 
\begin{equation}
 \Bigl\langle \log \sigma|M_r\Bigr\rangle_{\rm SDSS-B06} 
%        = (2.190\pm 0.080) - (0.100\pm 0.007)\,(M_{r}+21)
        = 2.190 - 0.100\,(M_{r}+21)
 \label{sdssVL}
\end{equation}
with intrinsic scatter of 0.07~dex; it happens that this is not 
very different from the relation obtained by Bernardi et al. (2003b).  

A common problem from which all local samples suffer is that, 
while the apparent magnitude of an object can be measured quite 
accurately,  the absolute magnitude is more difficult because 
it depends on the distance to the galaxy.  The true distance is 
difficult to measure because the redshift, which is usually 
well-determined, is a combination of the distance to the galaxy 
and the component of its peculiar velocity which is directed along 
the line of sight to the observer:  
 $cz = Hd + v_{\rm los}$.  
Typical velocities are expected to be of order a few hundred 
km~s$^{-1}$, so the redshift is a reliable distance indicator only 
beyond about $100h^{-1}$Mpc.  Most of the objects in the SDSS lie 
well beyond this distance, whereas all local samples are shallower.  
Thus, a legitimate concern is whether uncertainties in estimating 
the true distance are driving the difference between the SDSS 
measurement and those which are based on more local samples.   

To address such concerns, we have studied the $\sigma-L$ relation 
in the definitive sample of nearby early-type galaxies---that assembled 
in the ENEAR database (da Costa et al. 2000; Bernardi et al. 2002a;
Alonso et al. 2003; Wegner et al. 2003).  
ENEAR contains about 1000 objects out to 7000~km~s$^{-1}$, for which 
measured redshifts and estimated distances are available.  
%(Note that ENEAR is about five times larger than the Prugniel-Simien 
%sample.)  
%Using a spectroscopic sample of about 120,000 early-type galaxies
%extracted from the SDSS database (the 4th data release: 
%Adelman-McCarthy et al. 2006)
%we were able to match about 50 ENEAR objects with available imaging 
%and/or spectroscopy.  
Section~2 shows how the $\langle\log\sigma|\log L\rangle$ 
relation changes if one uses the redshift rather than the estimated 
distance when computing $L$.  
It highlights the fact that, if $\sigma$ played a role in determining 
the distance, e.g. if the distance comes from a Fundamental Plane or 
$D_n-\sigma$ analysis, then the $\sigma-L$ relation may be biased.  
%Section~2 also shows that the ENEAR and SDSS estimates 
%of $\sigma$ agree well, whereas the luminosities are rather different.  
In Section~3 we discuss our results.

\section{The local $\sigma-L$ relation}
The first part of this section compares various determinations of 
the local $\sigma-L$ relation.  The second part compares these 
determinations with that from the SDSS-B06.  

\begin{figure}
 \centering
 \includegraphics[width=0.9\columnwidth]{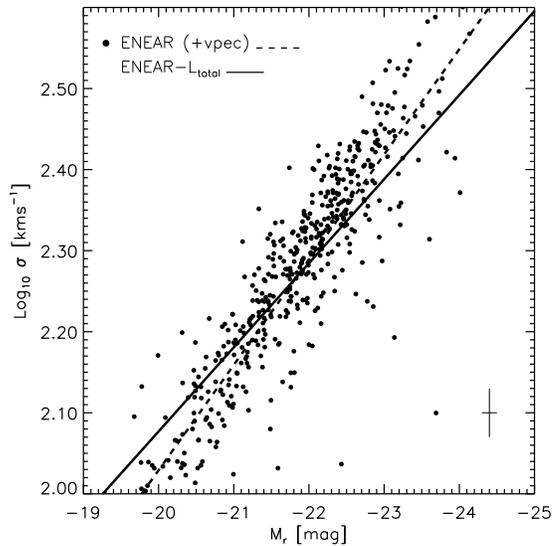}
 \caption{Joint distribution of $L$ and $\sigma$ in the ENEAR sample.  
   when $L$ has been estimated using a distance which has been 
   corrected for peculiar motions.  Short dashed line shows 
   $\langle\sigma|L_d\rangle$.  
   Using the redshift as distance indicator instead leads to the 
   $\langle\sigma|L_z\rangle$ relation shown as the solid line.
   The symbols and lines show the result of using total rather 
   than bulge luminosities for ENEAR.  }
 \label{localLV}
\end{figure}

\subsection{ENEAR}
The ENEAR sample (da~Costa et al. 2000) is approximately magnitude 
limited to 14.5 in the B band.  The catalog provides new apparent 
magnitudes, $d_n$ measurements, redshifts, and velocity dispersions 
for about 1000 early-type galaxies distributed over the whole sky 
out to about 7000~km~s$^{-1}$.  

In principle, estimating $\langle\sigma|L\rangle$ in the ENEAR 
sample is straightforward because the ENEAR team has also 
published a measurement of the $D_n-\sigma$ correlation:  
\begin{eqnarray}
 \log_{10} \left({D_n\over {\rm km~s}^{-1}}\right) &=& 1.406
  + 1.203 \log_{\rm 10}\left({\sigma\over {\rm km~s}^{-1}}\right)\nonumber\\
  &&     - \log_{10} \left({d_n\over 0.1~{\rm arcmin}}\right)
 \label{DnSigma}
\end{eqnarray}
(Bernardi et al. 2002b), with a scatter of 0.2/ln(10)~dex.  
Here, $d_n$ is that angular scale within which the average 
surface brightness in the R$_c$ band is 19.25~mag$/$arcsec$^2$.
For reasons we describe below, we compute luminosities for the 
ENEAR galaxies based on two different distance estimates.  
One uses this $D_n-\sigma$ relation to estimate `true' distances, 
and the other uses the redshift as a distance indicator (i.e., 
this second distance estimate ignores peculiar velocities).  
We will refer to the associated luminosities as $L_d$ and $L_z$.  
In addition, for ease of comparison with the SDSS, we shift all 
magnitudes from ENEAR $R_C$ to SDSS $r = 0.24 + R_C$.

\begin{figure*}
 \centering
\includegraphics[width=1.9\columnwidth]{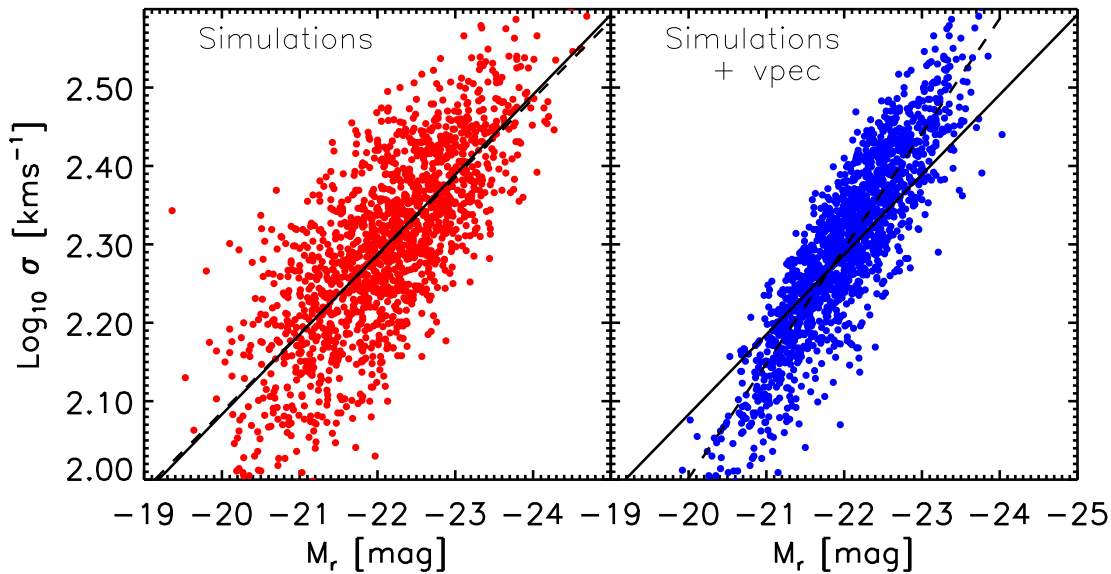}
 \caption{Joint distribution of $L$ and $\sigma$ in a mock catalog 
          in which the underlying galaxy population has been 
          constructed to mimic the SDSS-B06 sample, and to which 
          the ENEAR magnitude and redshift selection cuts have 
          been applied.  The two panels differ in how the luminosities 
          were estimated:  the panel on the left uses the observed 
          redshift to define a distance, whereas the panel on the 
          right makes a correction for peculiar velocities which 
          is based on the $D_n-\sigma$ relation.  In both panels, 
          solid line shows the input $\langle\sigma|L\rangle$ relation, 
          and dashed line shows the relation defined by the points in 
          the panel.  }
 \label{mockLV}
\end{figure*}

The joint distribution of $\sigma$ and $L_d$ which results from 
using the $D_n-\sigma$ distance estimate (filled circles) is 
shown in Figure~\ref{localLV}. Here we only show galaxies which have
both spectroscopy and photometry observed by the ENEAR team (i.e.
we did not include measurements listed in the ENEAR catalog which 
were compiled from previous work). In addition, we selected galaxies
with disk-to-bulge ratio smaller than 0.5 and used the total
magnitude listed by Alonso et al. (2003) (using the bulge luminosity 
the results only slightly change).
Dashed line shows 
\begin{equation}
 %\langle\log_{10}\sigma|M\rangle_{\rm ENEARd} = -0.127\,M_r - 0.457
 \Bigl\langle\log_{10}\sigma|M\Bigr\rangle_{\rm ENEARd} 
%  = 2.21 - 0.127\,(M_r + 21)
  = 2.159 - 0.130\,(M_r + 21)
 \label{eneard}
\end{equation}
which best-fits the sample.  Comparison with equation~(\ref{sdssVL}) 
shows that this relation is considerably steeper than in the SDSS.  

The observed scatter around this relation is $\sim 0.08$~dex.  This 
is about 0.01~dex larger than what SDSS reports as intrinsic scatter.  
This is remarkable, because ENEAR ought to be carrying distance errors 
of about twenty percent.  These errors ought to translate into 
increased scatter of 0.43~mags along the x-axis of 
Figure~\ref{localLV}.  If not accounted-for (and the fit above does 
not), this should have artificially decreased the slope and increased 
the scatter of the $\langle\sigma|L\rangle$ correlation.  
E.g., adding this scatter to the SDSS numbers would have decreased 
the magnitude of the slope by about fifteen percent, and the rms 
scatter around the new relation would have been about fifteen percent 
larger.

\subsection{To remove $v_{\rm pec}$ or not to remove $v_{\rm pec}$?}
Since the $D_n-\sigma$ estimate of the distance is only good to 
20 percent, one might well wonder if samples like ENEAR reach 
distances at which the redshift itself provides a more reliable 
estimate of the true distance than relations like $D_n-\sigma$ or 
the Fundamental Plane.  This will happen on scales $r$ where 
 $(v_{\rm pec}/300~{\rm km~s}^{-1})/0.2 < (r/3h^{-1}{\rm Mpc})$.  
This suggests that, beyond about $50h^{-1}$Mpc, the redshift may 
actually be a better estimate of the true distance even if typical 
line-of-sight peculiar velocities were as high as 1000~km~s$^{-1}$.  

Since ENEAR straddles this distance regime, we have studied what 
happens to the $\sigma-L$ relation when the redshift itself is used 
as an estimate of the distance---i.e., no correction is made for 
the peculiar velocity.  The inferred relation, when the total luminosity 
is used, 
\begin{equation}
 \Bigl\langle\log_{10}\sigma|M\Bigr\rangle_{\rm ENEARz} 
 = 2.184 - 0.104\,(M_r + 21),
 \label{enearztot}
\end{equation}
is shown as the dashed line in Figure~\ref{localLV}. The relation
obtained when the total luminosity is scaled to the bulge luminosity
\begin{equation}
 \Bigl\langle\log_{10}\sigma|M\Bigr\rangle_{\rm ENEARz-bulge} 
 = 2.198 - 0.102\,(M_r + 21),
 \label{enearzbulge}
\end{equation}
is shown as the dotted line. The two relations are similar.
Equation~\ref{enearztot} is 
considerably shallower than $\langle\sigma|L_d\rangle$ 
(equation~\ref{eneard}), with substantially larger scatter (0.12~dex).  
%A similar analysis of the PS sample results in the long dashed line 
%shown in Figure~\ref{localLV}; 
%\begin{equation}
% \langle\log_{10}\sigma|M\rangle_{\rm ENEARz} = -0.100\,(M_r+21) + 2.182,
% \label{psz}
%\end{equation}
%it too is shallower than when the distance indicator was corrected 
%for peculiar velocities.  

\begin{figure*}
 \centering
 \includegraphics[width=1.2\columnwidth]{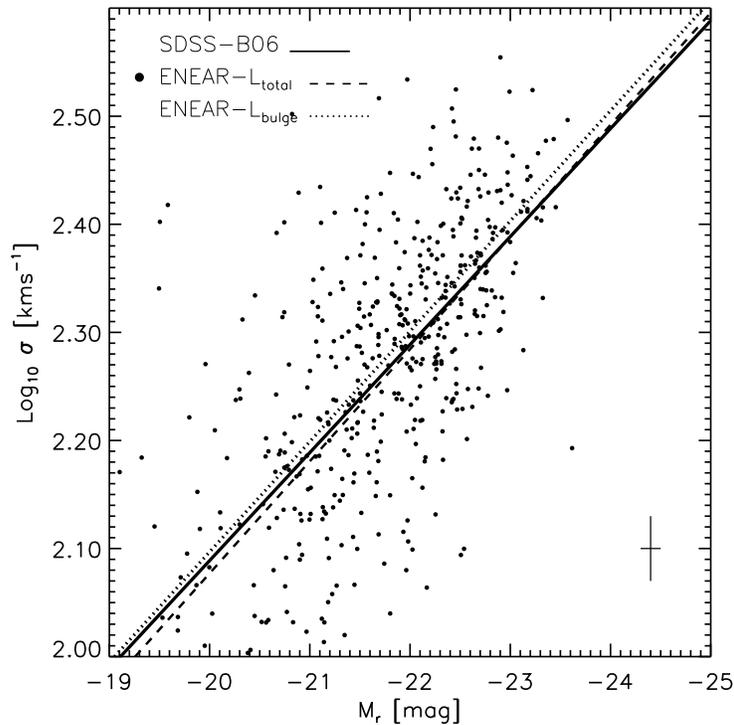}
 \caption{Joint distribution of $L$ and $\sigma$ in the ENEAR sample 
   %  and PS samples (filled and open symbols) 
      when the luminosity is computed using the redshift distance; 
      i.e., the distance has not been corrected for peculiar motions.
      Short dashed lines show the $\langle\sigma|L_z\rangle$ relation 
      defined by these points, and solid line shows the relation 
      defined by the SDSS-B06 sample.  
      Dotted line shows the result of using bulge rather than total 
      luminosities for ENEAR.}
 \label{localVLz}
\end{figure*}

Comparison of the relation based on redshift, $\sigma-L_z$ say, 
with the one based on distances, $\sigma-L_d$, shows that the 
scatter appears to decrease considerably when using the distances 
rather than the redshifts.  While it is tempting to conclude that 
this is signalling that the distance indicator is accurate, this is 
not the full story.  
After all, the $D_n-\sigma$ distance estimate is actually a function 
of $\sigma$:  distance~$\propto \sigma^{1.2}/d_n$.  
As a result, $\sigma$ appears in both axes of Figure~\ref{localLV}, 
with the x-axis proportional to $m - 5\log_{10}(\sigma^{1.2}/d_n)$.
Therefore, scatter in the relation will be correlated along a line 
which has slope $y = -5 (1.2x)$.  A slope of $-1/6$ is not far-off 
the one actually observed.  Hence, it may be that the relation shown 
in Figure~\ref{localLV} and quantified by equation~(\ref{eneard}) is 
both biased and artificially tight.  
Note that similar concerns would also apply to the use of 
Fundamental Plane-based distance estimates (such as those used 
by Pruniel \& Simien 1996), since the coefficient of the velocity 
dispersion in the direct fit is also $\sim 1.2$
(e.g. J{\o}rgensen et al. 1996; Bernardi et al. 2003c). 
This discussion provides another reason why the simple procedure 
of using the redshift as the distance indicator might actually be 
the preferred one.

\subsection{Results from a mock ENEAR-SDSS catalog}\label{mocks}
To better understand the effect of using a $\sigma$-based distance 
estimate on the $\sigma-L$ relation, we constructed a mock catalog 
of the ENEAR sample.  This was done by {\em assuming} that the 
joint distribution in luminosity, size and velocity dispersion 
for ENEAR galaxies is the same as for SDSS-B06 galaxies, so we could 
simply follow the steps described by Bernardi et al. (2003b).  
We then added Gaussian noise with dispersion equal to the ENEAR 
observational estimates, and applied the ENEAR apparent magnitude 
and redshift cuts.  

Since this method allows us to generate both $L$ and a half-light 
radius $R_e$, the assumption that the light profile is deVaucoleur 
allows us to compute a value of $d_n$ for each mock galaxy.  We then 
fit for the $D_n-\sigma$ relation, finding that it had slope 1.15, 
zero-point 1.4 and rms scatter 0.1~dex.  These values are rather 
similar to equation~(\ref{DnSigma}), suggesting that our mock 
catalog is actually rather realistic.

Figure~\ref{mockLV} compares the $\sigma-L_d$ and $\sigma-L_z$ 
relations in the mock catalog;  the relation based on the distance 
indicator is clearly steeper and tighter than the one based on the 
redshift, just as in the ENEAR sample itself.  Moreover, notice that 
the slope of $\langle\sigma|L_d\rangle$ relation is clearly steeper 
than the input slope, whereas $\langle\sigma|L_z\rangle$ is in 
rather good agreement with the input relation.  This suggests 
that equation~(\ref{enearztot}) is closer to the true relation for 
ENEAR galaxies than is equation~(\ref{eneard}).  
This bias does not depend strongly on the range of apparent magnitudes
in the sample. It does depend on the intrinsic distribution of the 
absolute magnitudes.
When the absolute magnitude range is large then the slope of the
$\sigma-L_d$ relation is closely related to that of the $D_n-\sigma$ relation 
(see discussion in previous subsection) and has little scatter.
However, when the luminosity function is narrow (i.e. sharply peaked)
then the $\sigma-L_d$ relation shows larger scatter: while the slope is 
still biased it is less easily related to that of the $D_n-\sigma$ relation.

%For similar reasons, we expect equation~(\ref{psd}) to have a 
%steeper slope than is actually true of the local 
%$\langle\sigma|L\rangle$ relation.  

\subsection{Comparison with SDSS-B06}
The analysis above suggests that the $\sigma-L_z$ relation in 
ENEAR, and, by extension, other local samples, are likely to 
be more reliable than the local $\sigma-L_d$ relations.  
Figure~\ref{localVLz} shows the joint distribution of $L_z$ and 
$\sigma$ in the ENEAR sample when 
the luminosity is computed using the redshift distance; i.e., the 
distance has not been corrected for peculiar motions.  The dashed 
line shows equation~(\ref{enearztot}), and the dotted line shows 
the result of using bulge rather than total luminosities for ENEAR.  
%The long dashed line shows the corresponding fit to the PS sample.  

These fits should be compared to the solid line, which shows the 
SDSS-B06 relation (equation~\ref{sdssVL}), for which the redshift is 
an excellent indicator of the true distance.  Note that the ENEAR 
and SDSS-B06 samples are in excellent agreement, suggesting that 
correcting for peculiar motions in local samples can lead to 
serious biases in correlations which involve $\sigma$.  
%Unfortunately, we have no explanation for why the PS sample is 
%so discrepant, although Figure~\ref{enearPS} suggests that 
%differences in photometric reductions are to blame.

\section{Discussion}
We showed that the ENEAR $\sigma-L_d$ relation, where $L_d$ indicates 
that the luminosity is based on a distance estimate which has been 
corrected for peculiar motions, is almost certainly biased.  This 
is because the distance indicator used to estimate $L_d$ depends on 
$\sigma$.  
We also showed that if the redshift is used as the distance 
indicator (i.e., no correction for peculiar motions is made) then 
the resulting $\sigma-L_z$ relation, while noisier, is less likely 
to be biased.  
%An object by object comparison showed that the ENEAR 
%velocity dispersions are in reasonably good agreement with those 
%from the Prugniel-Simien sample, provided that the PS values are 
%taken from HyperLeda.  The magnitudes show less agreement:  
%ENEAR tends to brighter by about 0.2~mags.  As a result, the PS 
%$\sigma-L_z$ relation is offset relative to the ENEAR relation.  
The ENEAR $\sigma-L_z$ relation is in excellent agreement with 
that measured in the SDSS, where neglecting peculiar velocities 
is an excellent approximation.  
Thus, it appears that the discrepancy between the SDSS $\sigma-L$ 
relation and that in local samples is due to the use of a 
distance indicator which correlates with $\sigma$.  

%The SDSS velocity dispersions are in good agreement with 
%those in ENEAR, the SDSS apparent magnitudes tend to be 
%slightly fainter than those in ENEAR, whereas the absolute magnitudes 
%are much more discrepant when $D_n-\sigma$ (or Fundamental Plane) 
%rather than redshift is used as the distance indicator.  

Our results are of particular interest for the problem of 
estimating black hole abundances from the observed distribution 
of luminosities or velocity dispersions 
(e.g. Yu \& Tremaine 2002; Shankar et al. 2004; 
Tundo et al. 2006).  The first method requires 
$\langle M_\bullet|L\rangle$ whereas the second requires 
$\langle M_\bullet|\sigma\rangle$.  The two approaches will only 
give the same estimate of $\phi(M_\bullet)$ if the 
$\langle \sigma|L\rangle$ relation of the black hole sample is 
the same as that of the sample from which the luminosity function
$\phi(L)$ and velocity dispersion function $\phi(\sigma)$ are drawn.  
Current black hole samples are relatively local, so their 
$\sigma-L$ correlations are very different from 
that of the SDSS. Bernardi et al. (2006b) suggests that current 
black hole samples are biased towards objects with abnormally large 
velocity dispersions for their luminosities. If this is a selection 
rather  than physical effect, then the $M_\bullet-\sigma$ and $M_\bullet-L$ 
relations currently in the literature are also biased from their 
true values. Bernardi et al. find that the bias in the 
$\langle M_\bullet|\sigma\rangle$ relation is likely to be small, 
whereas the $\langle M_\bullet|L\rangle$ relation is biased towards 
predicting more massive black holes for a given luminosity.
%Our results suggest that this is probably because 
%it is difficult to estimate $L$ reliably in local samples.  
Therefore, the estimate based on the velocity function (e.g. that of 
Sheth et al. 2003) is to be preferred.  

The $\langle\sigma|L\rangle$ relations show evidence for a 
flattening at large $L$, a fact we do not use here, but which is 
relevant to studies of BCGs (e.g. Bernardi et al. 2006a) and may 
be relevant to studies of black hole demographics.  

\acknowledgements

I would like to thank Ravi Sheth for discussions on how to generate
the ENEAR mock catalog and Monique Aller for comparisons of 
velocity dispersion measurements.
This work is partially supported by NASA grant LTSA-NNG06GC19G, 
and by grants 10199 and 10488 from the Space Telescope Science 
Institute, which is operated by AURA, Inc., under NASA contract 
NAS 5-26555.

Funding for the SDSS and SDSS-II has been provided by the 
Alfred P. Sloan Foundation, the Participating Institutions, 
the NSF, the US DOE, NASA, 
the Japanese Monbukagakusho, the Max Planck Society 
and the Higher Education Funding Council for England.  
The SDSS website is http://www.sdss.org/.

The SDSS is managed by the Astrophysical Research Consortium (ARC) 
for the Participating Institutions:  The American Museum of Natural 
History, Astrophysical Institute Postdam, the University of Basel, 
Cambridge University, Case Western Reserve University, 
the University of Chicago, Drexel University, Fermilab, 
the Institute for Advanced Study, the Japan Participation 
Group, the Johns Hopkins University, the Joint Institute for 
Nuclear Astrophysics, the Kavli Institute for Particle Astrophysics 
and Cosmology, the Korean Scientist Group, the Chinese Academy 
of Sciences (LAMOST), Los Alamos National Laboratory, 
the Max Planck Institute for Astronomy (MPI-A), 
the Max Planck Institute for Astrophysics (MPA), 
New Mexico State University, the Ohio State University, 
the University of Pittsburgh, the University of Portsmouth, 
Princeton University, the U.S. Naval Observatory, and the 
University of Washington.

\appendix
\section{Comparison of the $\sigma-L$ relation in various SDSS data releases}
Bernardi et al. (2003b) report the first SDSS-based $\sigma-L$ relation, 
from a sample of about $9000$ objects classified as being early-types 
on the basis of imaging (concentrated light profiles) and spectroscopy 
(weak or no emission lines):  hereafter, we will refer to this 
sample as SDSS-B03.  

\subsection{Changes to SDSS photometry}
Shortly after this estimate was published, the SDSS reported a problem 
with the point-spread-function in the SDSS photometry which led, on average, 
to an overestimate of 
the apparent brightness of extended objects and larger effective radii.  
This problem was fixed in subsequent SDSS data releases.  
However, more recent work has shown that there remain a problem with 
the SDSS photometric reductions which is most severe for bright 
objects in crowded fields (the SDSS pipeline overestimates the sky level; 
see Mandelbaum et al. 2005; Bernardi et al. 2006a; Lauer et al. 2006).  
Re-analysis of such objects (Bernardi et al. 2006a; Hyde et al. 2006) 
suggests that the SDSS-DR5 photometric reductions 
underestimate the true apparent brightness by 0.1~mags on average, 
and by up to 0.5~mags for bright objects in crowded fields  
(by chance the combination of the two problems make the magnitudes 
in Bernardi et al. 2003a to be similar to the recomputed values
by Hyde et al.).

These changes in the photometric reductions affect the estimated $L$ 
in the $\sigma-L$ relation.  They also have a small effect on $\sigma$, 
because the velocity dispersions are scaled (to account for aperture 
effects) to a fraction (typically 1/8) of the half light radius, and 
the half light radii are about 15\% smaller than in 
Bernardi et al. (2003a).  This aperture correction is small, so the 
change to the $\sigma-L$ relation is driven by the change to $L$.  

\begin{figure}[b]
 \centering
 \includegraphics[width=0.9\columnwidth]{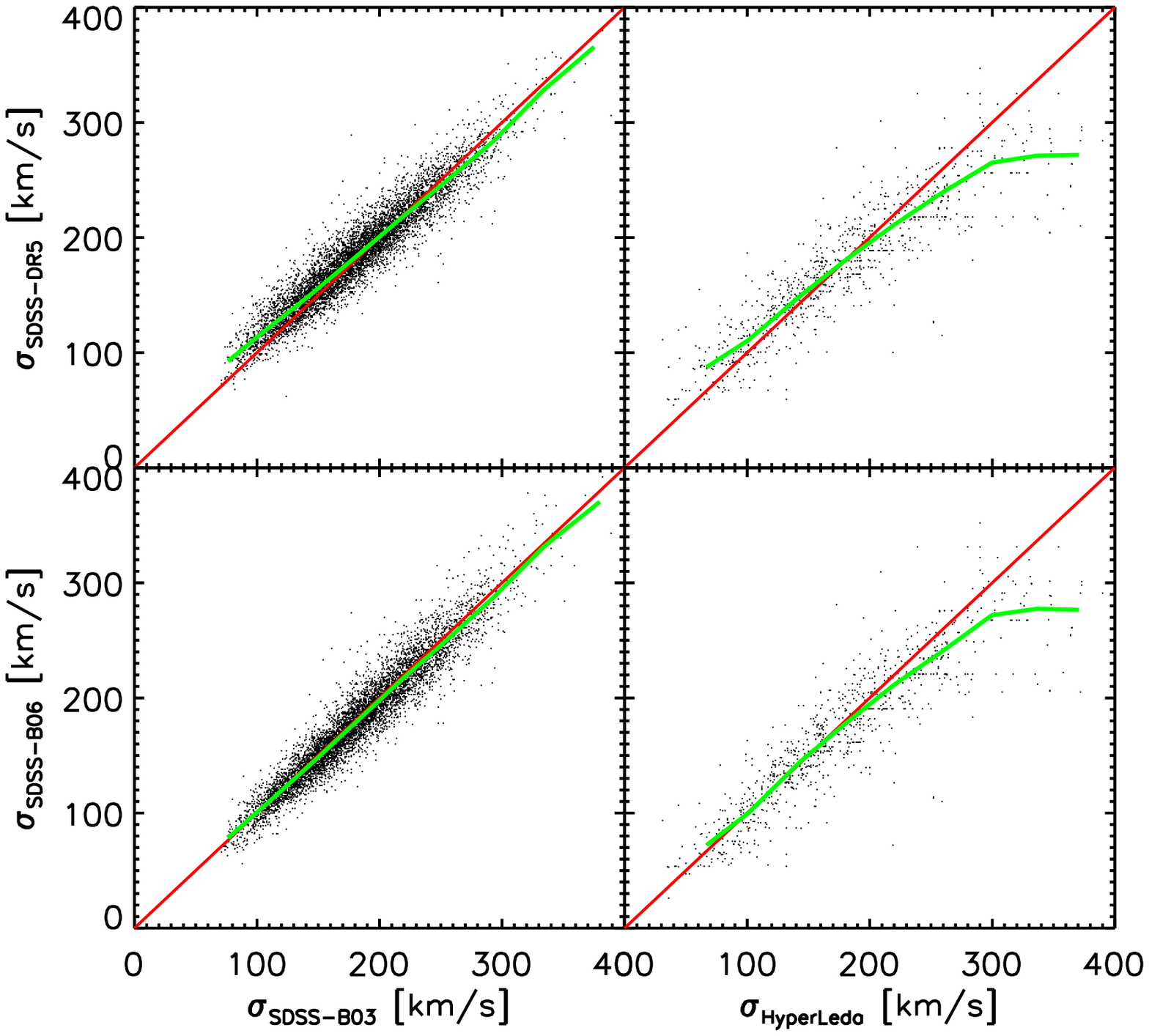}
 \caption{Comparison of $\sigma$ values from Bernardi et al. (2003a) 
     (left panels) and HyperLeda (right panels) with those from 
     SDSS-DR5 (top) and with the values used in this paper 
     (bottom).  
     Top left:  At fixed SDSS-B03, the SDSS-DR5 values are systematically 
     higher when $\sigma\le 150$~km~s${-1}$.  
     Bottom left:  The same, but after correcting for this bias by 
     re-analyzing the spectra.
     Top right:  At small $\sigma$, SDSS-DR5 reports larger values than 
     HyperLeda; the offset is similar to that seen in the left-hand panel; 
     in this case, we know that SDSS-DR5 is biased high.
     HyperLeda reports substantially larger velocity dispersions 
     at $\sigma\ge 250$~km~s$^{-1}$.  The following figures suggest 
     that, in this case, SDSS is more reliable than HyperLeda.
     Bottom right:  The same, but after correcting for this bias by 
     re-analyzing the spectra.     }
 \label{sdssB03}
\end{figure}

\begin{figure}
 \centering
 \includegraphics[width=0.6\columnwidth]{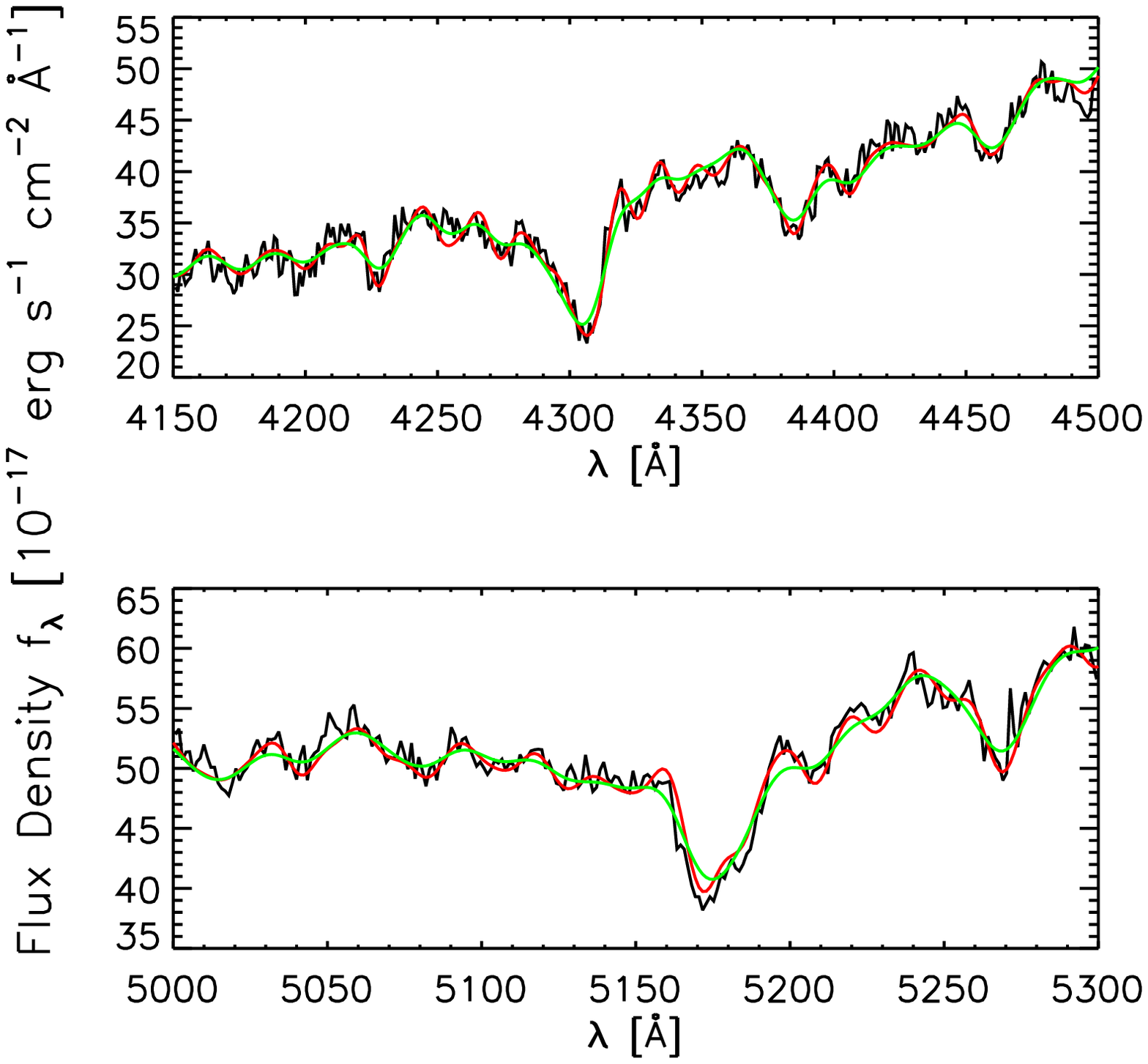}
 \caption{Comparison of the SDSS observed spectrum of an object in common
    with the literature (jagged line) with templates broadened by the velocity 
    dispersions reported by SDSS-DR5 (288~km~s$^{-1}$) and 
    HyperLeda (436~km~s$^{-1}$) (red and green lines respectively).  
    Top and bottom panels show two sections of the spectrum.  
    The smaller velocity dispersion is clearly a better fit.}
 \label{testEFAR1}
\end{figure}

\begin{figure}
 \centering
 \includegraphics[width=0.45\columnwidth]{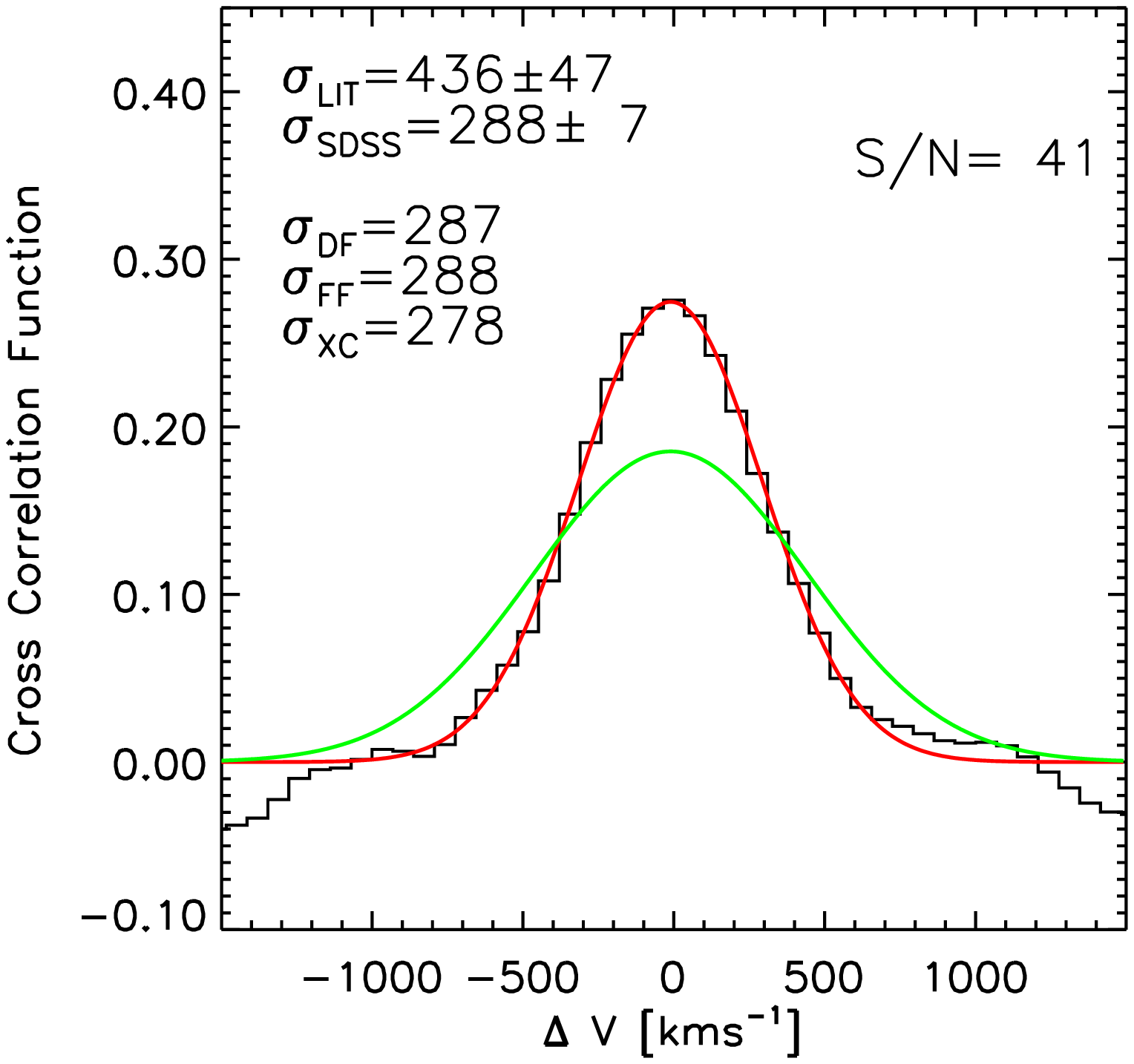}
 \caption{Cross correlation peak for the object shown in the 
    previous Figure.  The SDSS-DR5 value is clearly a better 
    description of this broadening function (red lines was computed
    using the SDSS velocity dispersion while green is from the value in the
    literature).  Panel also shows estimates of the velocity dispersion 
    obtained from the {\it Direct Fitting}, {\it Fourier Fitting} and 
    {\it cross-correlation} methods:  the SDSS-DR5 pipeline returns the 
    average of the first two methods.  }
 \label{testEFAR2}
\end{figure}

\subsection{Changes to SDSS spectroscopy}
In addition, the SDSS spectroscopic reductions appear to have 
changed between Bernardi et al. (2003a) and DR5.  The left-hand panel of 
The top left panel in Figure~\ref{sdssB03} shows that $\sigma$ in 
the SDSS-DR5 for the SDSS-B03 sample do not match the values used 
by Bernardi et al. (2003a).
The difference is small but systematic, with SDSS-DR5 being 
larger than SDSS-B03 at small $\sigma$.  
Since there is considerably more overlap between measurements from 
the literature (from the HyperLeda database) with SDSS-DR5 than with SDSS-B03,
we have checked if the SDSS-DR5 velocity dispersion values are biased
at low $\sigma$.   
The top right hand panel of Figure~\ref{sdssB03} shows that 
SDSS-DR5 is biased towards larger $\sigma$ at $\sigma\le 150$~km~s$^{-1}$;
The bias is similar to that seen on the top left panel.  

The SDSS-DR5 values are based on averaging the measurements from the
{\it Direct-Fitting} and {\it Fourier-fitting} methods (Bernardi et al. 2003a).
We have run simulations similar to those in Bernardi et al. (2003a) 
and found that the discrepancy results from the fact that the
{\it Fourier-fitting} method is now biased $~15\%$ level at low sigma 
($\sim 100 kms^{-1}$), whereas the other method is not.  
The bottom panels of Figures~\ref{sdssB03}
shows that using only the {\it Direct-Fitting} method improves the 
agreement between SDSS-DR5 and SDSS-B03 as well as between SDSS-DR5 
and HyperLeda.  

The right panels in Figure~\ref{sdssB03} show that HyperLeda reports 
substantially larger velocity dispersions at $\sigma\ge 250$~km~s$^{-1}$.  
Figures~\ref{testEFAR1} and~\ref{testEFAR2} suggest that, in this 
case, SDSS is more reliable than HyperLeda (these figures are a 
representative example; we have checked all galaxies with velocity 
dispersion larger than $250$~km~s$^{-1}$).  
This is important, because the objects with the largest $\sigma$ 
are expected to host the most massive black holes.  
If the SDSS determinations were 
indeed biased low, then determinations of black hole abundances 
which are based on the SDSS velocity function (Sheth et al. 2003) 
would underestimate the abundances of supermassive black holes 
(e.g. McClure \& Dunlop 2004;  Shankar et al. 2004; 
Tundo et al. 2006). Our analysis shows that this is not a concern.

\subsection{The SDSS-B06 sample and the $\sigma-L$ relation}
Because of the problems with the SDSS photometry and spectroscopy, we 
have chosen to use our new estimate of the $\sigma-L$ relation.  
We selected the SDSS-B03 sample of early-type galaxies.
For the photometry, we use the reductions from Hyde et al. (2006).  
For the velocity dispersions, we use the {\it Direct-Fitting} method 
described above, but do not average it with the {\it Fourier-fitting} 
method (as done for the values in the SDSS database).  
In the main text, we refer to the catalog which results as the 
SDSS-B06 sample.

\begin{figure}
 \centering
 \includegraphics[width=0.45\columnwidth]{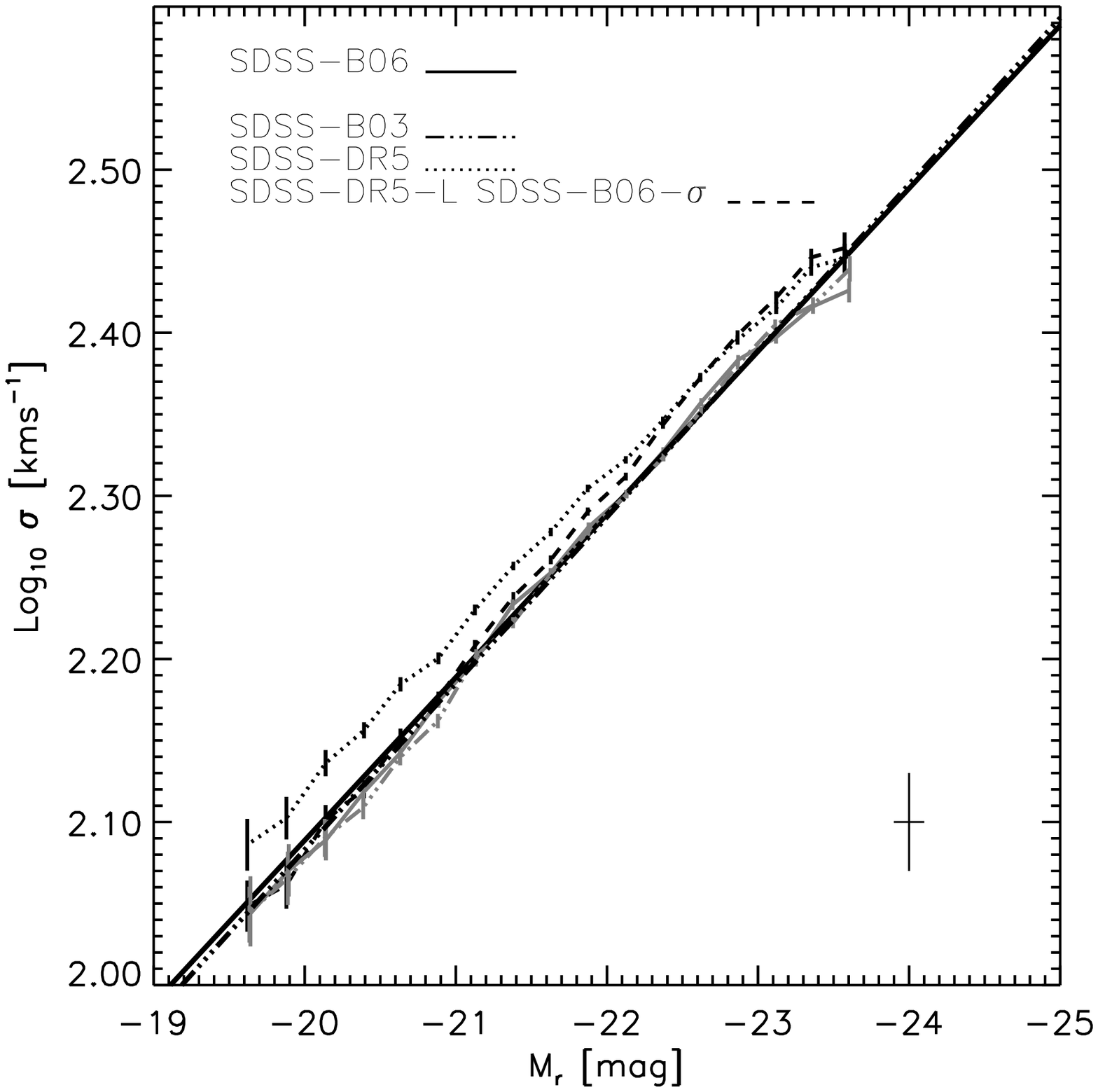}
 \includegraphics[width=0.45\columnwidth]{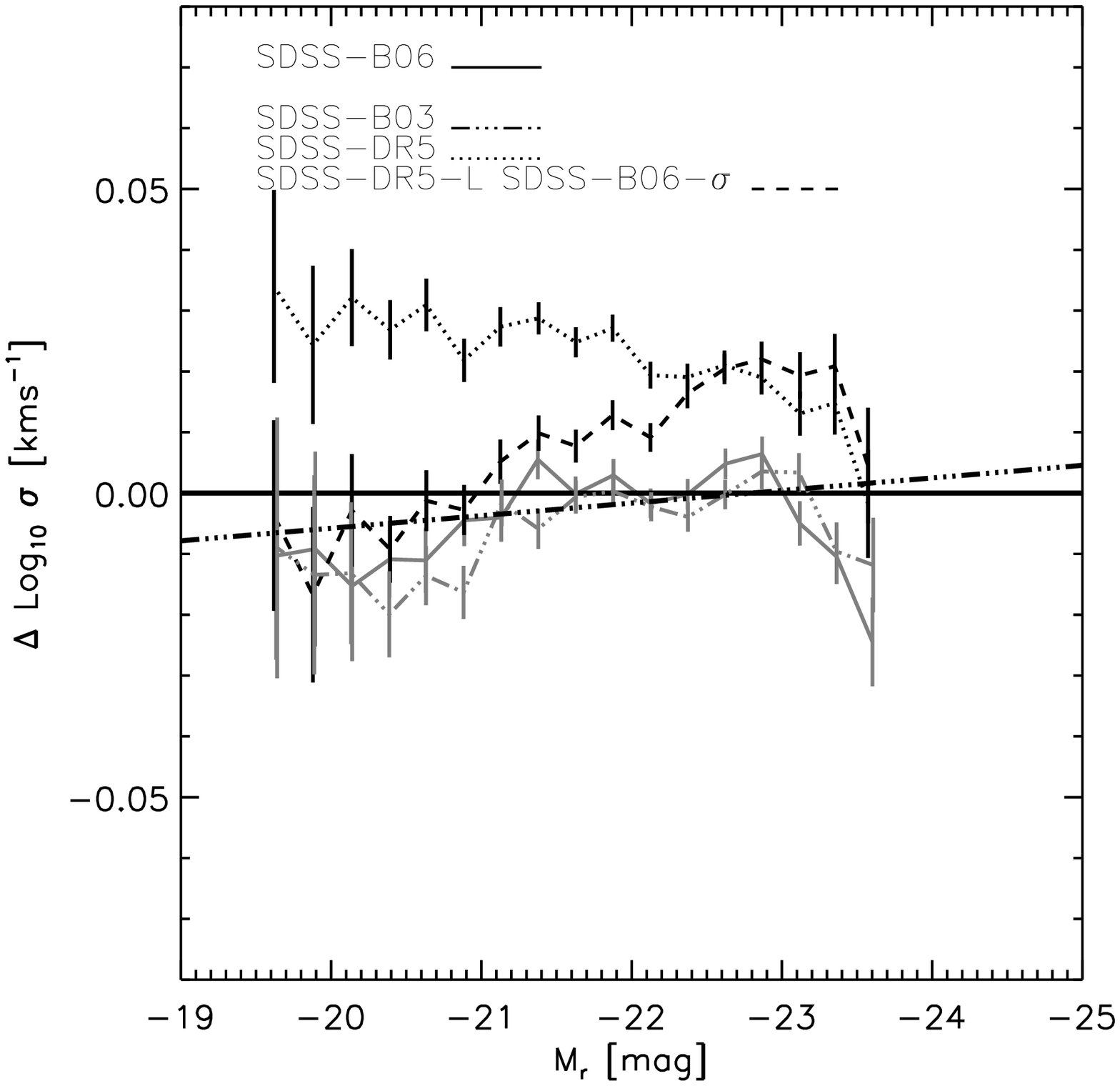}
 \caption{Comparison of $\sigma-L$ relation in various SDSS reductions.  
          Panel on the left shows the relations themselves, and panel 
          on the right shows the relations expressed as residuals 
          from the fit given in Equation~(\ref{sdssVL}).  Although this 
          is not the main point of the current study, note that the 
          relation clearly flattens at large magnitudes.}
 \label{sdssInter}
\end{figure}

The main text studies the $\sigma-L$ relation. 
Figure~\ref{sdssInter} illustrates the effect of these various SDSS 
reductions on this relation.  In all cases, straight lines show 
single power-law fits to $\langle\sigma|L\rangle$, and jagged lines 
show the median $\sigma$ as a function of $L$ for a few bins in $L$.  
To highlight the differences, the y-axis in the left hand panel shows 
 $\sigma - 2.19 - 0.100(M_r+21)$, rather than $\sigma$ itself, as a 
function of $L$.  
The triple-dot dashed lines show the relation reported by 
Bernardi et al. (2003b), and the dotted lines show fits based on 
quantities output by SDSS-DR5 (corrected for aperture effects to 
$R_e/8$).  The short dashed lines are based on SDSS-DR5 photometry, 
but use the SDSS-B06 dispersions (to approximately 
correct for the small systematic bias in SDSS-DR5 $\sigma$s), and solid 
lines show the result of recomputing the photometry as well as the 
velocity dispersions (the SDSS-B06 sample).  

%The dotted line, which is based on SDSS DR5 quantities, is most 
%discrepant from the others; it has a shallower slope, and is similar 
%to the relation reported by Mitchell et al. (2005) on the basis of 
%the Bernardi et al. (2003) sample after the official changes to the 
%SDSS photometry and spectroscopy had been made
%\begin{equation}
% \langle\log_{10}\sigma|M_r\rangle_{\rm SDSS} = -0.092\,(M_r+21) + 2.22.
% \label{sdss05}
%\end{equation}
The relations from SDSS-B03 and SDSS-B06 are rather similar 
(the various adjustments to the SDSS-B03 photometry approximately cancel).  
Notice that all these relations show evidence for a flattening at 
large $L$, a fact we do not use in the main text, but which is 
relevant to studies of BCGs (e.g. Bernardi et al. 2006a) and may 
be relevant to studies of black hole demographics.

\subsection{Object by object comparison with ENEAR}
\label{sdss-enear}

%We argued that most of the discrepancy between the ENEAR and SDSS 
%$\sigma-L$ relations is due to the use of the $D_n-\sigma$ distance 
%indicator when estimating $L$.  Here we provide other evidence that 
%this must be the case.  
%Using a spectroscopic sample of about 120,000 early-type galaxies
%extracted from the SDSS database (the 4th data release: 
%Adelman-McCarthy et al. 2006)

The main text compares the $\sigma-L$ relation in the SDSS-B06 
and ENEAR samples.  In the remainder of this section we provide 
an object by object comparison of the two sets of photometric and 
spectroscopic reductions.  We began by matching the two catalogs:  
only about 50 objects with ENEAR imaging and/or spectroscopy were 
also in the SDSS-DR5 sample.  
This surprisingly small number is because the ENEAR objects are, 
in general, very bright and very large compared to the vast 
majority of objects targeted for SDSS spectroscopy.  
Of these, 30 have ENEAR velocity dispersions and 29 have ENEAR 
photometry, but only 15 have both.  

\begin{figure}
 \centering
\includegraphics[width=0.45\columnwidth]{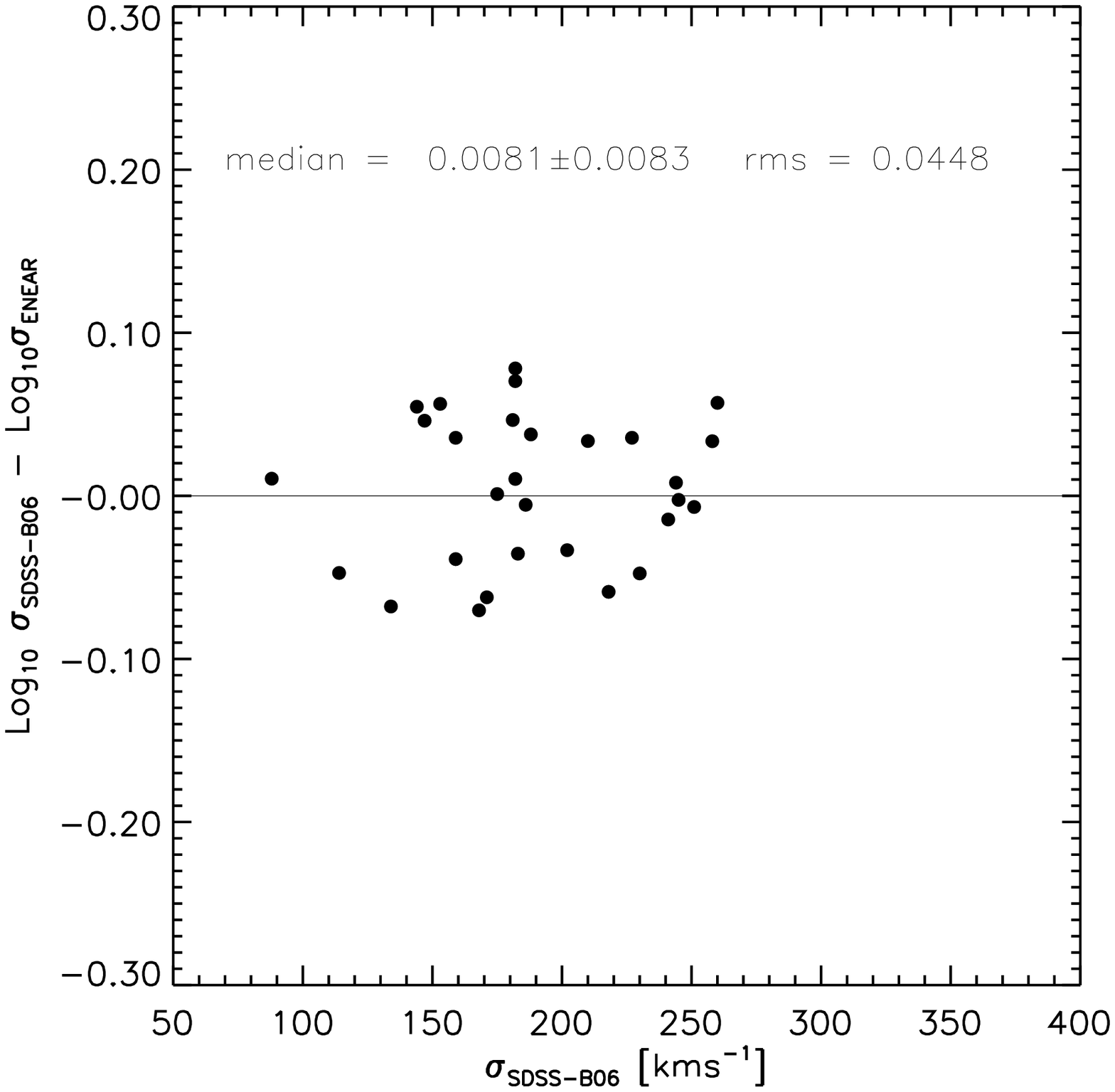}
\includegraphics[width=0.45\columnwidth]{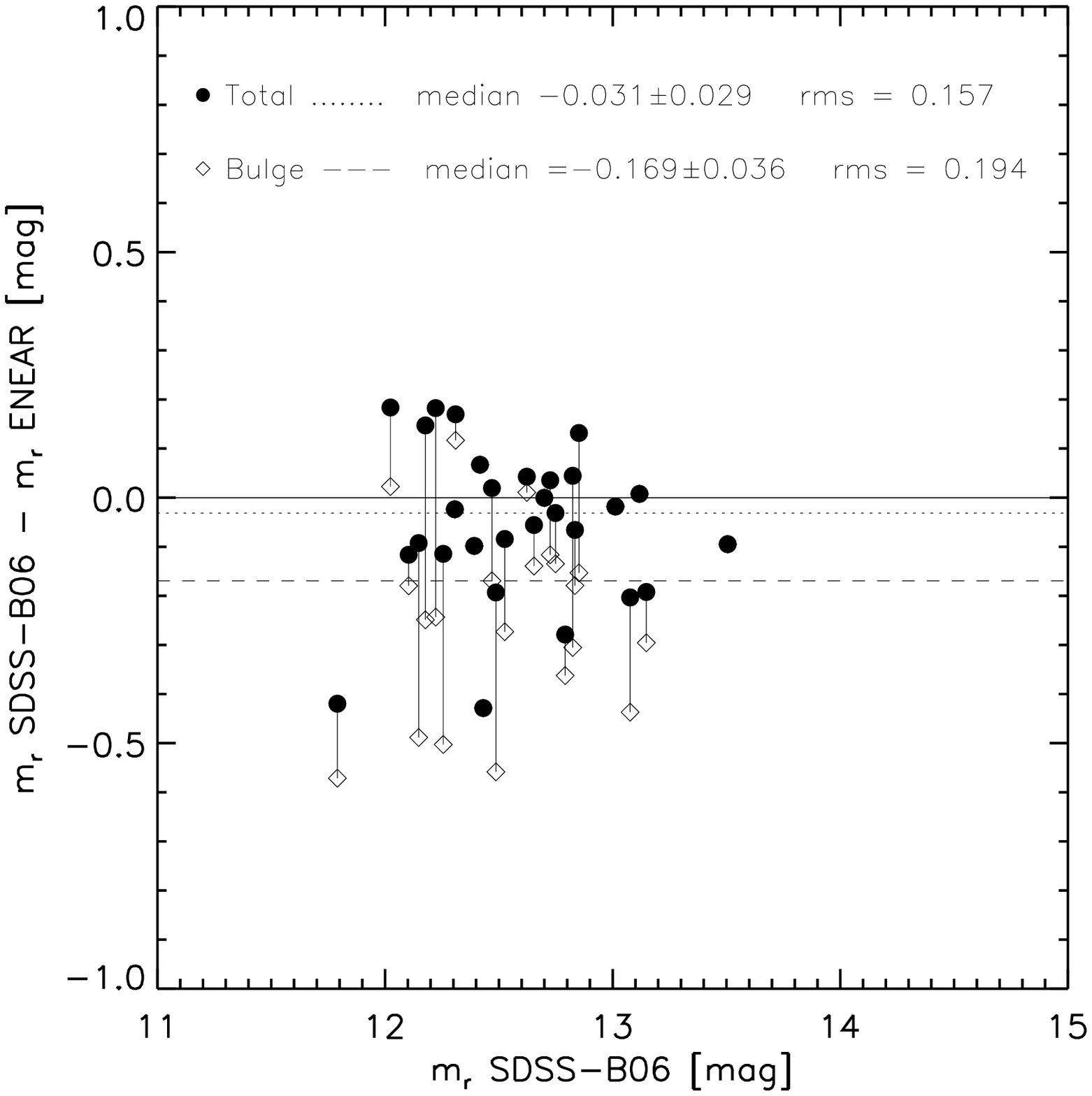}
 \caption{Left: Comparison of ENEAR and {\it Direct-Fitting} estimates of the 
          velocity dispersions of the 30 objects in common.  There is 
          no offset, and the magnitude of the scatter is consistent 
          with being entirely due to observational errors.  
          Right: Comparison of ENEAR and Hyde et al. estimates of the 
          apparent magnitudes of the 29 objects in common.  Whereas 
          Hyde et al. reports a magnitude associated with fitting a 
          single deVaucoleur profile to the image, ENEAR reports 
          both bulge (open diamonds) and total (filled circles)
          magnitudes.  On average, the Hyde et al. magnitude is similar 
          to ENEAR total magnitude, but is 0.12~mags brighter
          than ENEAR bulge magnitude.  }
 \label{vsdss=venear}
 \label{msdss=benear}
\end{figure}

Figure~\ref{vsdss=venear} compares the ENEAR and the {\it Direct-Fitting}
estimates of the velocity dispersions of the 30 objects which have 
ENEAR spectroscopy available.  The estimates are generally in good 
agreement, suggesting that SDSS-B06 velocity dispersions are unlikely 
to be systematically biased with respect to local samples.  
Therefore, the discrepancy in the $\sigma-L$ relation is almost 
certainly due to systematic errors in the luminosity.  

Such errors could arise from systematic differences in the 
photometry itself, or in the conversion from apparent to absolute 
magnitude.  To test this, Figure~\ref{msdss=benear} compares the 
ENEAR and Hyde et al. estimates of the apparent magnitudes for 29 galaxies
which have ENEAR imaging available. Whereas Hyde et al. reports a 
magnitude associated with fitting a single deVaucoleur profile to 
the image, ENEAR reports both total (filled circles) and bulge 
(open diamonds) magnitudes.  
The figure indicates that Hyde et al. magnitudes tend to be similar
to the ENEAR total magnitudes, and about 0.12~mags brighter 
than ENEAR bulge magnitudes. Here we compare galaxies dominated
by the bulge component (i.e. with disk-to-bulge ratio less than 0.5).  

%The SDSS is known to underestimate the luminosities of bright 
%galaxies in crowded fields (Lauer et al. 2006; Bernardi et al. 2006a;
%Hyde et al. 2006) or bright nearby galaxies due to sky problems, 
%so the discrepancies shown here are not unexpected.  Note, however, 
%that these differences are smaller than the $\sim 0.5$~mag 
%discrepancy seen at $M<-23$ in Figure~\ref{localLV}, and they have 
%the wrong sign;  correcting for this bias would increase the SDSS 
%magnitudes, exacerbating the discrepancy in the $\sigma-L_d$ relation
%while decreasing the difference between the SDSS and ENEAR $\sigma-L_z$
%relations.  

\begin{figure}
 \centering
\includegraphics[width=0.45\columnwidth]{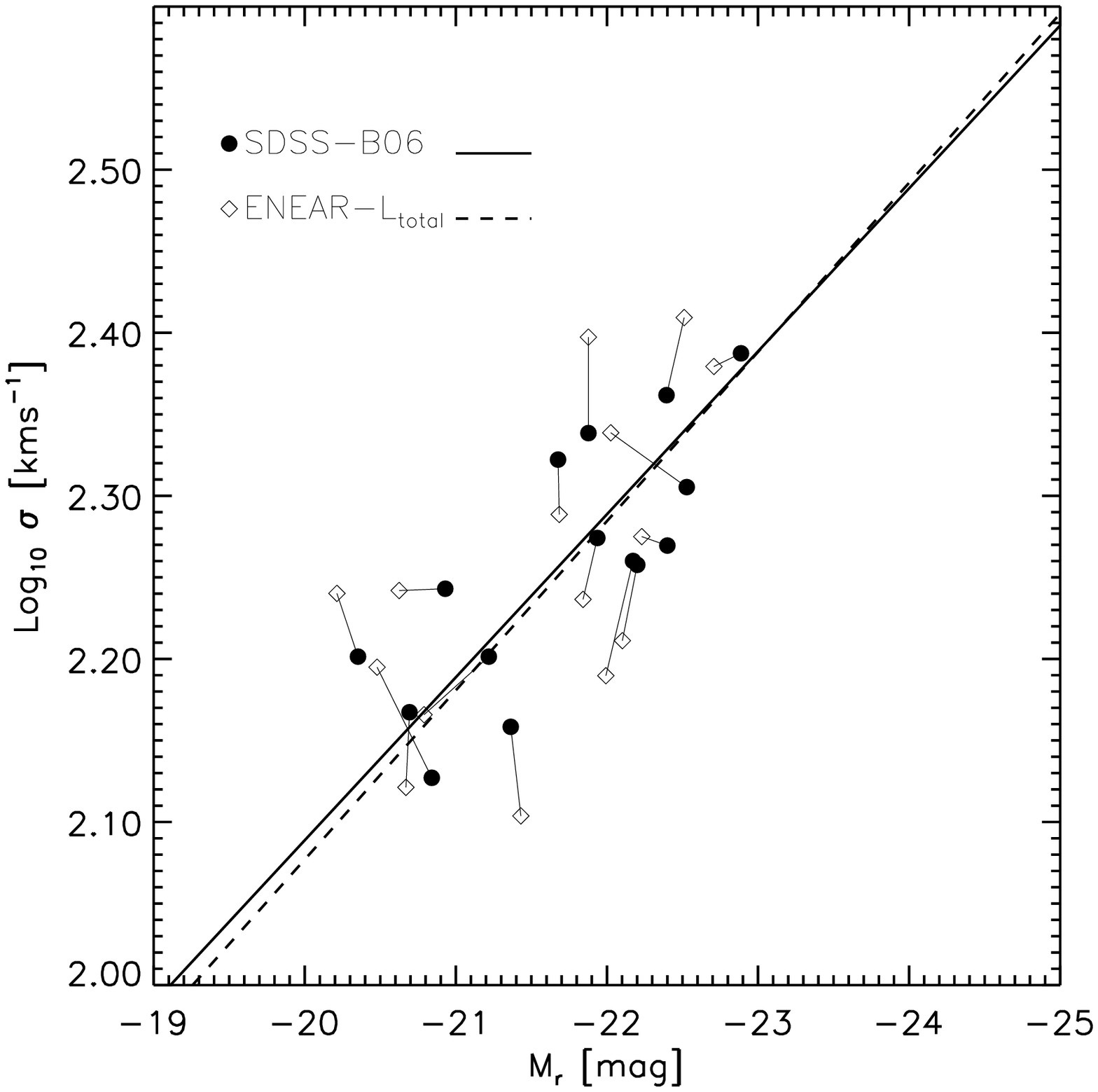}
 \caption{Joint distribution of ENEAR (open diamonds) and the new recomputed 
          SDSS (filled circles) luminosity and velocity dispersion estimates 
          for the 15 objects common to both surveys with available 
          imaging and spectroscopy.  Dashed and solid lines show 
          $\langle\sigma|L\rangle$ for the full ENEAR and SDSS-B06 
          samples, when the luminosities are estimated from the 
          redshifts (i.e. no correction for peculiar velocities is made).  }
 \label{commonLV}
\end{figure}

Figure~\ref{commonLV} shows the $\sigma-L_z$ relation for objects 
with both ENEAR imaging and spectroscopy in common with SDSS.  
It is in reasonably good agreement with the SDSS-B06 relation.

\end{document}